\documentstyle[12pt,cite,epsf]{article}

\newcommand{\aleq}{\mathrel
  {\raisebox{-1ex}{$\stackrel{\textstyle <}{\sim}$}}}

\begin{document}

\title{Curvature and scaling in 4D dynamical triangulation}

\author{Bas V. de Bakker\thanks{email: bas@phys.uva.nl} \\Jan
  Smit\thanks{email: jsmit@phys.uva.nl} \and Institute for Theoretical
  Physics, University of Amsterdam\\ Valckenierstraat 65, 1018 XE
  Amsterdam, the Netherlands.}

\date{12 December 1994}

\maketitle

\begin{abstract}
  We study the average number of simplices $N'(r)$ at geodesic
  distance $r$ in the dynamical triangulation model of euclidean
  quantum gravity in four dimensions. We use $N'(r)$ to explore
  definitions of curvature and of effective global dimension.  An
  effective curvature $R_V$ goes from negative values for low
  $\kappa_2$ (the inverse bare Newton constant) to slightly positive
  values around the transition $\kappa_2^c$. Far above the transition
  $R_V$ is hard to compute. This $R_V$ depends on the distance scale
  involved and we therefore investigate a similar explicitly $r$
  dependent `running' curvature $R_{\rm eff}(r)$.  This increases from
  values of order $R_V$ at intermediate distances to very high values
  at short distances. A global dimension $d$ goes from high values in
  the region with low $\kappa_2$ to $d=2$ at high $\kappa_2$. At the
  transition $d$ is consistent with 4. We present evidence for scaling
  of $N'(r)$ and introduce a scaling dimension $d_s$ which turns out
  to be approximately 4 in both weak and strong coupling regions. We
  discuss possible implications of the results, the emergence of
  classical euclidean spacetime and a possible `triviality' of the
  theory.
\end{abstract}

{%
\thispagestyle{myheadings}
\renewcommand{\thepage}{ITFA-94-23}

\clearpage
}

\section{Introduction}

The dynamical triangulation model \cite{AgMi92a,AgMi92b,AmJu92} is a
very interesting candidate for a nonperturbative formulation of
four-dimensional euclidean quantum gravity.  The configurations in the
model are obtained by glueing together equilateral four-dimensional
simplices in all possible ways such that a simplicial manifold is
obtained.  A formulation using hypercubes was pioneered in
\cite{We82}.  The simplicial model with spherical topology is defined
as a sum over triangulations ${\cal T}$ with the topology of the
hypersphere $S^4$ where all the edges have the same length $\ell$. The
partition function of this model is
\begin{equation}
Z(N,\kappa_2) = \sum_{{\cal T},\,N_4=N} \exp(\kappa_2 N_2).
\label{partn}
\end{equation}
Here $N_i$ is the number of simplices of dimension $i$ in the
triangulation ${\cal T}$.  The weight $\exp(\kappa_2 N_2)$ is part of
the Regge form of the Einstein-Hilbert action $-\int \sqrt{g}
\penalty0 R\slash 16\pi G_0$,
\begin{eqnarray}
  -S & = & \frac{1}{16\pi G_0} \sum_{\Delta} V_{2}\, 2\delta_{\Delta} =
  \kappa_2 (N_2 - \rho N_4),
  \label{RE}
  \\ 
  \kappa_2 &=& \frac{V_2}{8G_0},\;\;\;
  \rho  =  \frac{10\arccos(1/4)}{2\pi} = 2.098 \cdots,
  \label{rho}
\end{eqnarray}
Here $V_2$ is the volume of 2-simplices (triangles $\Delta$) and
$\delta_{\Delta}$ is the deficit angle around $\Delta$.  The volume of
an $i$-simplex is
\begin{equation}
  V_i = \ell^i \sqrt{(i+1)/2^i}/i!.
  \label{simvolume}
\end{equation}
Because the $N_i$ $(i=0\ldots4)$ satisfy three constraints only two of
them are independent.  We have chosen $N_2$ and $N_4$ as the
independent variables.  For comparison with other work we remark that
if $N_0$ is chosen instead of $N_2$ then the corresponding coupling
constant $\kappa_0$ is related to $\kappa_2$ by $\kappa_0 = 2
\kappa_2$.  This follows from the relations between the $N_i$, which
imply that 
\begin{equation}
  N_0 - \frac{1}{2} N_2 + N_4 = \chi,
  \label{nrel}
\end{equation}
where $\chi$ is the Euler number of the manifold, which is 2 for
$S^4$.

Average values corresponding to (\ref{partn}) can be estimated by 
Monte Carlo methods, which require varying $N_4$ 
\cite{AgMi92a,AgMi92b,AmJu92}. One way to implement the condition 
$N_4=N$ is to base the simulation on the partition function 
\cite{AgMi92a} 
\begin{equation}
  Z'(N,\kappa_2) = \sum_{\cal T} 
  \exp(\kappa_2 N_2 - \kappa_4 N_4 - \gamma (N_4 - N)^2),
  \label{partv}
\end{equation}
where $\kappa_4$ is chosen such that $\langle N_4\rangle \approx N$
and the parameter $\gamma$ controls the volume fluctuations. The
precise values of these parameters are irrelevant if the desired $N_4$
are picked from the ensemble described by (\ref{partv}). This is not
done in practice but the results are insensitive to reasonable
variations in $\gamma$.  We have chosen the parameter $\gamma$ to
be $5 \cdot 10^{-4}$, giving $\langle N_4^2\rangle-\langle
N_4\rangle^2 \approx (2\gamma)^{-1} = 1000$, i.e.\ the
fluctuations in $N_4$ are approximately 30.

Numerical simulations
\cite{AgMi92a,AgMi92b,AmJu92,AmJuKr93,Var93,Br93,CaKoRe94a} have shown
that the system (\ref{partv}) can be in two phases\footnote{In
  section~\ref{discussion} we discuss the possibility that these are
  not phases in the sense of conventional statistical mechanics.}. For
$\kappa_2 > \kappa_2^c(N_4)$ (weak bare coupling $G_0$) the system is
in an elongated phase with high $\langle \overline{R}\rangle$, where
$\langle \overline{R}\rangle$ is the average Regge curvature
\begin{equation}
  \overline{R}= \frac{2\pi V_2}{V_4}(\frac{N_2}{N_4}-\rho)
  \longleftrightarrow \frac{\int \sqrt{g} R}{\int \sqrt{g}}.
\label{R0}
\end{equation}
In this phase the system has relatively large baby universes 
\cite{AmJaJuKr93} and resembles a branched polymer.
{} For $\kappa_2 < \kappa_2^c(N_4)$ (strong coupling) the
system is in a crumpled phase with low $\langle \overline{R}\rangle$.
This phase is highly connected, i.e.\ the average number of simplices
around a point is very large. The transition between the phases appears to be 
continuous.

\begin{figure}[t]
\epsfxsize=\textwidth
\epsffile{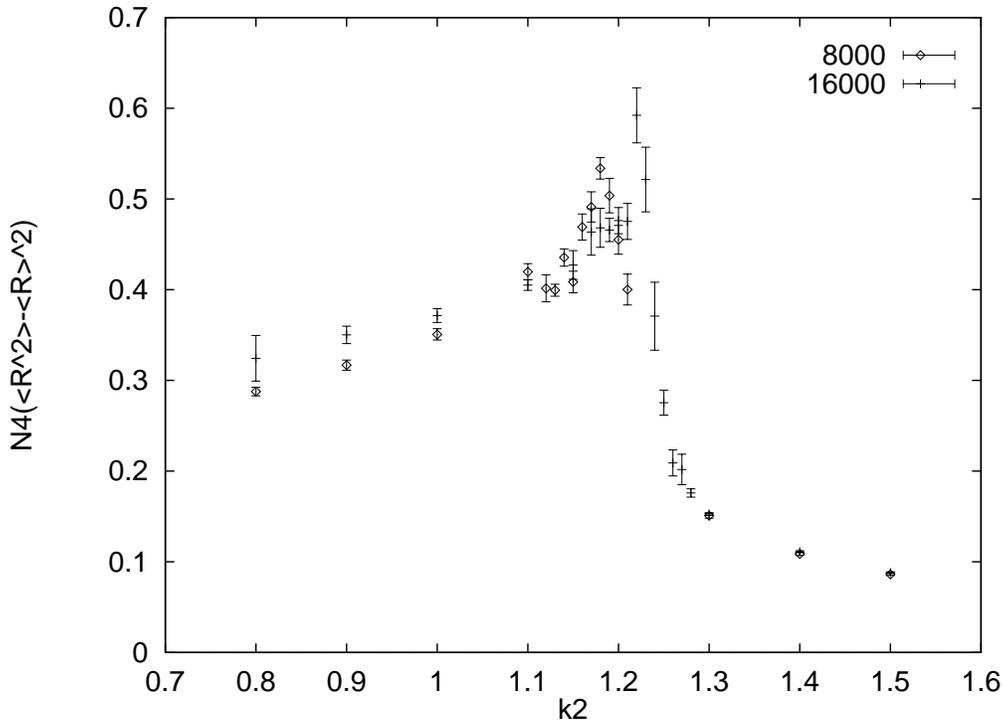}
\caption{The curvature susceptibility as a function of $\kappa_2$ for
  8000 and 16000 simplices.}
\label{suscepfig}
\end{figure}

As an example and for later reference we show in
figure~\ref{suscepfig} the susceptibility
\begin{equation}
  \left[\left\langle \left(\frac{N_2}{N_4}\right)^2 \right\rangle -
    \left\langle \frac{N_2}{N_4} \right\rangle^2 \right] N =
      \frac{V_4}{(2\pi V_2)^2} \left[ \langle \overline{R}^2 \rangle -
      \langle \overline{R} \rangle^2 \right] V.
\label{susc}
\end{equation}
($V=N_4 V_4$).  The two curves are for $N=8000$ and 16000 simplices.

The data for $N=8000$ are consistent with those published in reference
\cite{CaKoRe94a}, where results are given for
\begin{equation}
  \chi(N_0) = \frac{1}{N} \left( \langle N_0^2 \rangle - \langle N_0
  \rangle^2 \right).
\end{equation}
For fixed $N_4$ this is $1/4$ of our curvature susceptibility
(\ref{susc}), as can be seen from (\ref{nrel}).

The behavior of $Z(\kappa_2,N)$ as a function of $N$ for large $N$ has
been the subject of recent investigations
\cite{CaKoRe94b,AmJu94,BaSm94b,BruMa94}.  In ref.\ \cite{BaSm94b} we
discussed the possibility that $\kappa_2^c$ might move to infinity as
$N\rightarrow \infty$ and argued that this need not invalidate the
model. So far a finite limit is favoured by the data
\cite{AmJu94,BruMa94}, however.

It is of course desirable to get a good understanding of the
properties of the euclidean spacetimes described by the probability
distribution $\exp(-S)$.  A very interesting aspect is the
proliferation of baby universes \cite{AmJaJuKr93}.  Here we study more
classical aspects like curvature and dimension, extending previous
work in this direction
\cite{AgMi92a,AgMi92b,AmJu92,AmJuKr93,Var93,Br93,CaKoRe94a}.  The
basic observable for this purpose is the average number of simplices
at a given geodesic distance from the arbitrary origin, $N'(r)$.  We
want to see if this quantity can be characterized, approximately, by
classical properties like curvature and dimension, and if for suitable
bare couplings $\kappa_2$ there is a regime of distances where the
volume-distance relation $N'(r)$ can be given a classical
interpretation. It is of course crucial for such a continuum
interpretation of $N'(r)$ that it scales in an appropriate way.

In section~\ref{curvature} we investigate the properties of an
effective curvature for a distance scale that is large compared to the
basic unit but small compared to global distances.  Effective
dimensions for global distances are the subject of
section~\ref{dimension} and scaling is investigated in
section~\ref{scaling}.  We summarize our results in
section~\ref{summary} and discuss the possible implications in
section~\ref{discussion}.

\section{Curvature}
\label{curvature}

\begin{table}[t]
\newcommand{\tw}{3em}
\hfill
\begin{tabular}{|c|c|c|}
\hline
\makebox[\tw]{$\kappa_2$} & \makebox[\tw]{8000} & \makebox[\tw]{16000}
\\ \hline
0.80 & 29 & 16 \\ \hline
0.90 & 34 & 19 \\ \hline
1.00 & 38 & 25 \\ \hline
1.10 & 37 & 38 \\ \hline
1.12 & 13 & -- \\ \hline
1.13 & 24 & -- \\ \hline
1.14 & 28 & -- \\ \hline
1.15 & 21 & 13 \\ \hline
1.16 & 36 & -- \\ \hline
1.17 & 44 & 16 \\ \hline
1.18 & 90 & \phantom{0}8 \\ \hline
1.19 & 26 & 22 \\ \hline
\end{tabular}
\hfill
\begin{tabular}{|c|c|c|}
\hline
\makebox[\tw]{$\kappa_2$} & \makebox[\tw]{8000} & \makebox[\tw]{16000}
\\ \hline
1.20 & 56 & 45 \\ \hline
1.21 & 22 & 56 \\ \hline
1.22 & -- & 51 \\ \hline
1.23 & -- & 65 \\ \hline
1.24 & -- & 58 \\ \hline
1.25 & -- & 45 \\ \hline
1.26 & -- & 38 \\ \hline
1.27 & -- & 31 \\ \hline
1.28 & -- & 43 \\ \hline
1.30 & 43 & 41 \\ \hline
1.40 & 40 & 24 \\ \hline
1.50 & 47 & 21 \\ \hline
\end{tabular}
\hfill\hbox{}
\caption{Number of configurations used in the various calculations.}
\label{numtable}
\end{table}

A straightforward measure of the average local curvature is the bare
curvature at the triangles, i.e.\ $\langle \overline{R}\rangle =
48\pi\sqrt{3/5}\,\ell^{-2}\langle N_2/N_4 -\rho\rangle$ (this
follows from (\ref{R0}) and (\ref{simvolume})).  It has been well
established that at the phase transition $\langle N_2/N_4 - \rho
\rangle \approx 2.38 - 2.10 = 0.28$, practically independent of
the volume \cite{AgMi92a,AgMi92b,AmJu92,AmJuKr93}.  This means that
$\langle \overline{R}\rangle \approx 55 \ell^{-2}$ has to be divergent
in the continuum limit $\ell\rightarrow 0$.  However, the curvature at
scales large compared to the lattice distance $\ell$ is not
necessarily related to the average curvature at the triangles.  One
could imagine e.g.\ a spacetime with highly curved baby universes
which is flat at large scales.

The scalar curvature at a point is related to the volume of a small
hypersphere around that point.  Expanding the volume in terms of the
radius of the hypersphere results in the relation
\begin{eqnarray}
  V(r) &=& C_n r^n (1 - \frac{R_V \,r^2}{6(n+2)} + O(r^4)), 
\label{vofcur}\\
   C_n &=& \frac{\pi^{n/2}}{\Gamma(n/2+1)},
\end{eqnarray}
for an $n$-dimensional manifold.  We have written $R_V $ here to
distinguish it from the small scale curvature at the triangles.
Differentiating with respect to $r$ results in
the volume of a shell at distance $r$ with width $dr$,
\begin{equation}
  V'(r) = C_n n r^{n-1} (1 - \frac{R_V \,r^2}{6n} + O(r^4)).
  \label{vpofcur}
\end{equation}
 
We explore this definition of curvature as follows. We take the
dimension $n=4$, assuming that there is no need for a fractional
dimension differing from 4 at small scales. For $r$ we take the
geodesic distance between the simplices, that is the lowest number of
hops from four-simplex to neighbour needed to get from one
four-simplex to the other.  Setting the distance between the centers
of neighbouring simplices to 1 corresponds to taking a fixed edge
length in the simplicial complex of $\sqrt{10}$, i.e.\ we will use
lattice units with $\ell=\sqrt{10}$. For $V(r)$ and $V'(r)$ we take
\begin{eqnarray}
  V(r) = V_{\mbox{\scriptsize eff}} N(r),\;\;\; V'(r) =
  V_{\mbox{\scriptsize eff}} N'(r),\;\;\;N'(r) = N(r)-N(r-1) ,
  \label{VN}
\end{eqnarray}
where $N(r)$ is the average number of four-simplices within distance
$r$ from the (arbitrary) origin, $N'(r)$ is the number of simplices at
distance $r$ and we have allowed for an effective volume
$V_{\mbox{\scriptsize eff}}$ per simplex which is different from
$V_4$.  Since $R_V $ is to be a long distance (in lattice units)
observable we shall call it the effective curvature.

The effective curvature was determined by fitting the function
$N'(r)$ to
\begin{equation}
  N'(r)= ar^3 + br^5.
  \label{mcfit}
\end{equation}
It then follows from (\ref{vpofcur}) and (\ref{VN}) that
$R_V $ and $V_{\mbox{\scriptsize eff}}$ are determined by
\begin{equation}
  R_V  = -24 \frac{b}{a},\;\;\; V_{\mbox{\scriptsize eff}} = \frac{4C_4}{a}.
  \label{meanrfit}
\end{equation}
The constant $C_4$ in equation (\ref{vofcur}) is $\pi^2/2$.
In flat space we would have $V_{\mbox{\scriptsize eff}} = V_4$,
giving
\begin{equation}
  a = \frac{4 C_4}{V_4} = \frac{48 \sqrt{5} \pi^2}{125} \approx 8.47,
  \label{analytica}
\end{equation}
where we used $V_4 = 25\sqrt{5}/24$ for $\ell = \sqrt{10}$.  Such a
space cannot be formed from equilateral simplices because we cannot
fit an integer number of simplices in an angle of $2\pi$ around a
triangle.  We expect therefore $V_{\mbox{\scriptsize eff}}$ to be
different from $V_4$.

\begin{figure}[t]
\epsfxsize=\textwidth
\epsffile{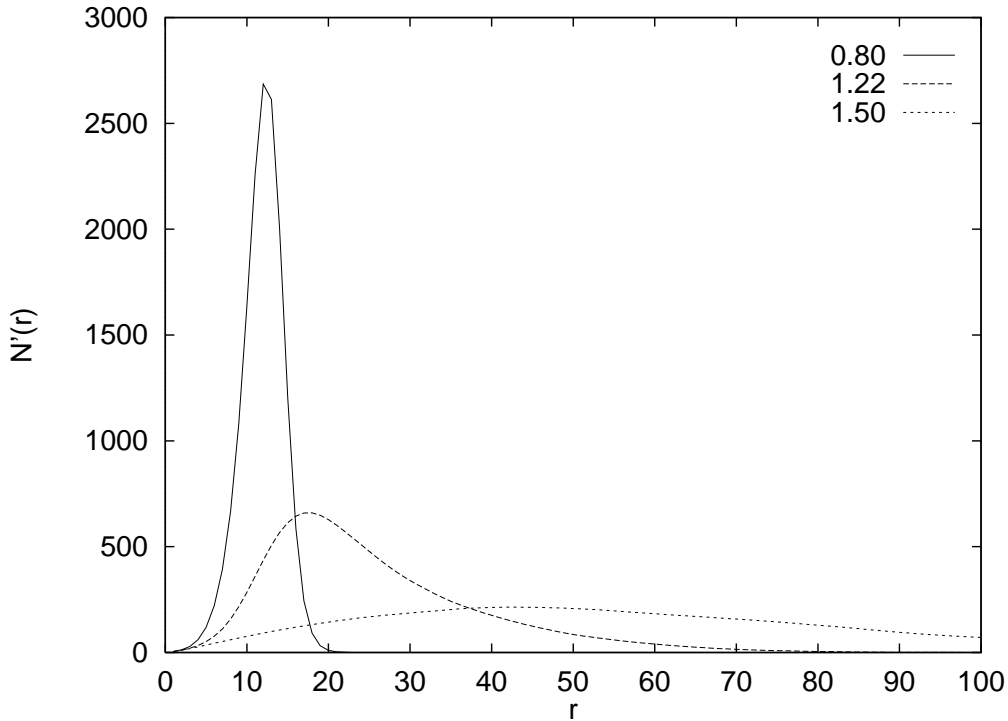}
\caption{The number of simplices $N'(r)$ at distance $r$ from the
  origin at $\kappa_2 = 0.80$, 1.22 and 1.50, for $N=16000$.}
\label{dvofrdrfig}
\end{figure}

Figure~\ref{dvofrdrfig} shows $N'(r)$ for 16000 simplices.  Three
different values of $\kappa_2$ are shown, 0.8 (in the crumpled phase),
1.22 (close to the transition) and 1.5 (in the elongated phase). These
curve can also be interpreted as the probability distribution of the
geodesic length between two simplices. Such distributions were
previously presented in refs.\ \cite{AmJu92,AmJuKr93}.

We determined $N'(r)$ by first averaging this value per configuration
successively using each simplex as the origin. We then used a
jackknife method, leaving out one configuration each time, to
determine the error in $R_V$.

Our results in this paper are based on $N=8000$ and $16000$ simplices.
Configurations were recorded every 10000 sweeps, where a sweep is
defined as $N$ accepted moves.  The time before the first
configuration was recorded was also 10000 sweeps.  We estimated the
autocorrelation time in the average distance between two simplices to
be roughly 2000 sweeps for $N=16000$ and $\kappa_2 = 1.22$.  For other
values of $\kappa_2$ the autocorrelation time was lower.  The number
of configurations at the values of $\kappa_2$ used are shown in
table~\ref{numtable}.

\begin{figure}[t]
\epsfxsize=\textwidth
\epsffile{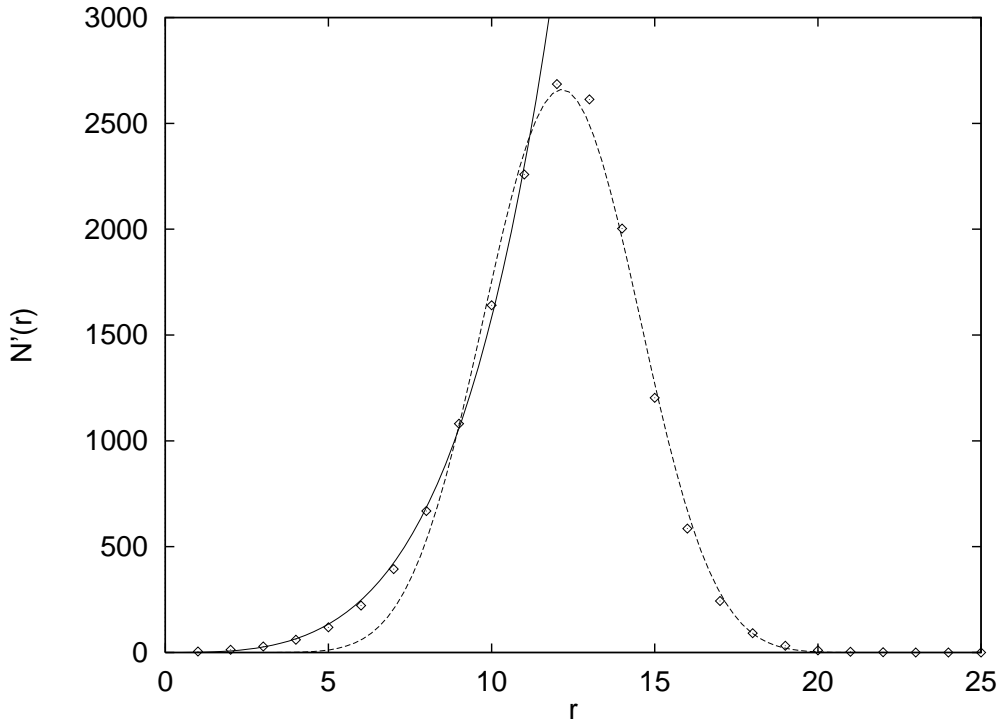}
\caption{Effective curvature fit and global dimension fit in the 
  crumpled phase at $\kappa_2=0.80$.}
\label{fitcru}
\end{figure}

\begin{figure}[t]
\epsfxsize=\textwidth
\epsffile{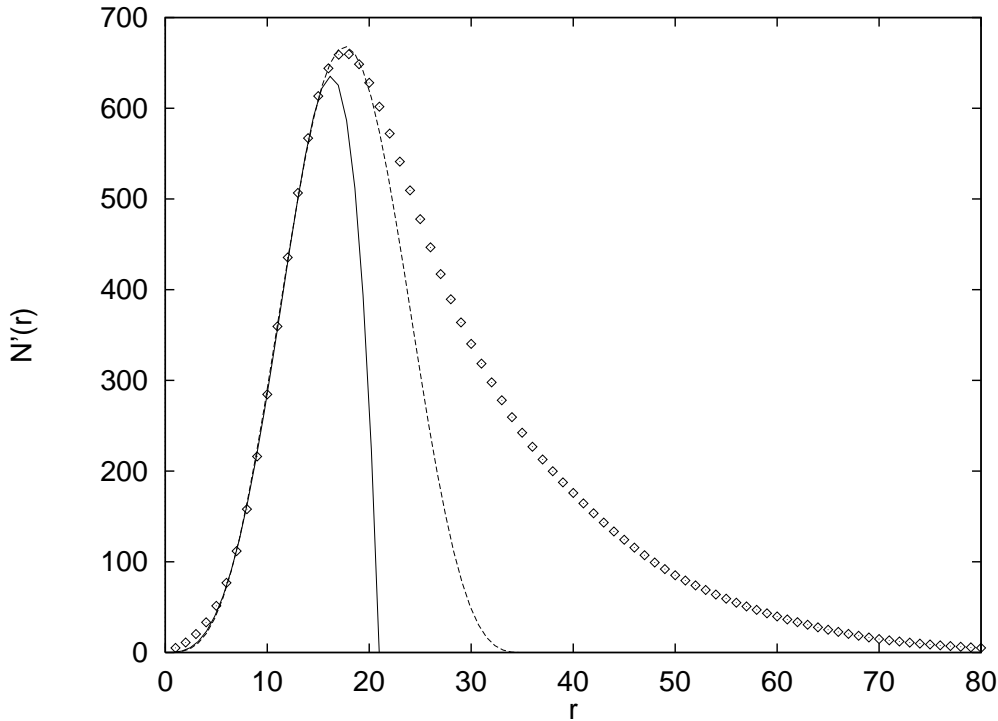}
\caption{Effective curvature fit and global dimension fit near the 
  transition, $\kappa_2=1.22$.}
\label{fittra}
\end{figure}

\begin{figure}[t]
\epsfxsize=\textwidth
\epsffile{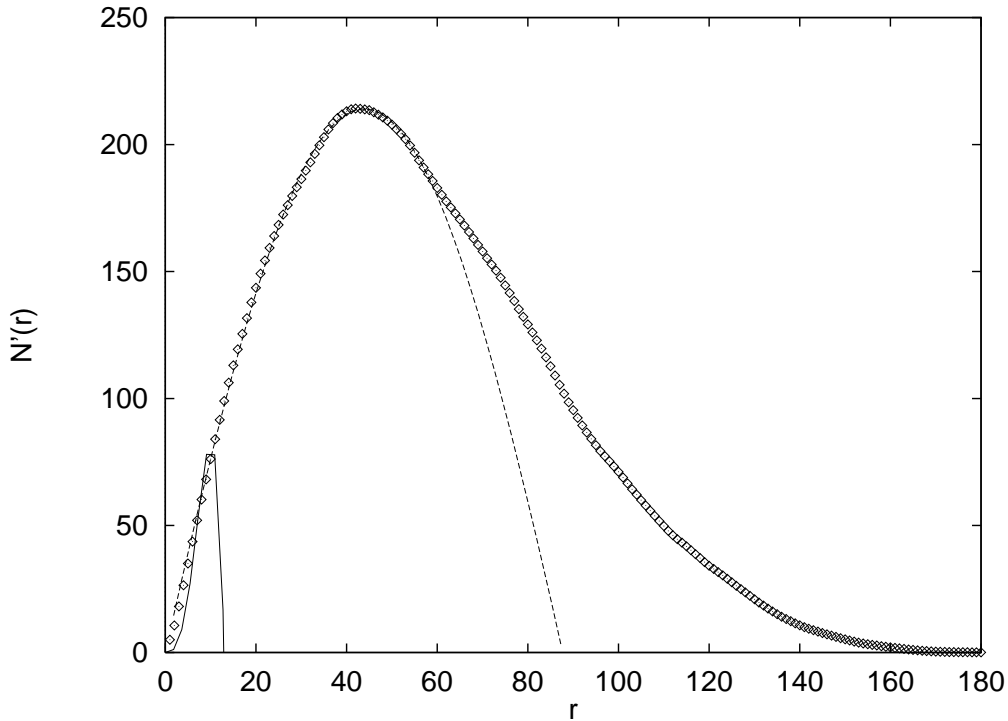}
\caption{Effective curvature fit and global dimension fit in the 
  elongated phase at $\kappa_2=1.50$ (the effective curvature fit is
  not appropriate here).}
\label{fitelo}
\end{figure}

{}Figures~\ref{fitcru}--\ref{fitelo} show effective curvature fits
(continuous lines) in the crumpled phase ($\kappa_2=0.80$), near the
transition ($\kappa_2=1.22$) and in the elongated phase
($\kappa_2=1.50$), together with global dimension fits (see the next
section) at longer distances, for $N=16000$.  The curves are extended
beyond the fitted data range, $r=1$--11, to indicate their region of
validity.  A least squares fit was used, which suppresses the lattice
artefact region where $N'(r)$ is small because it is sensitive to
absolute errors rather than relative errors.

For $\kappa_2 \aleq \kappa_2^c$ the fits were good even beyond the
range of $r$ used to determine the fit (except obviously when this
range already included all the points up to the maximum of $N'(r)$).
This can be seen in figure~\ref{fittra}, where the fit is good up to
$r = 15$.  The fit with $ar^3 + b r^5$ does not appear sensible
anymore for $\kappa_2$ values somewhat larger than the critical value
$\kappa_2^c$, because then $N'(r)$ goes roughly linear with $r$ down
to small distances.  This can be seen quite clearly in
figure~\ref{fitelo} (so the $R_V$ fit in this figure should be
ignored).

\begin{figure}[t]
\epsfxsize=\textwidth
\epsffile{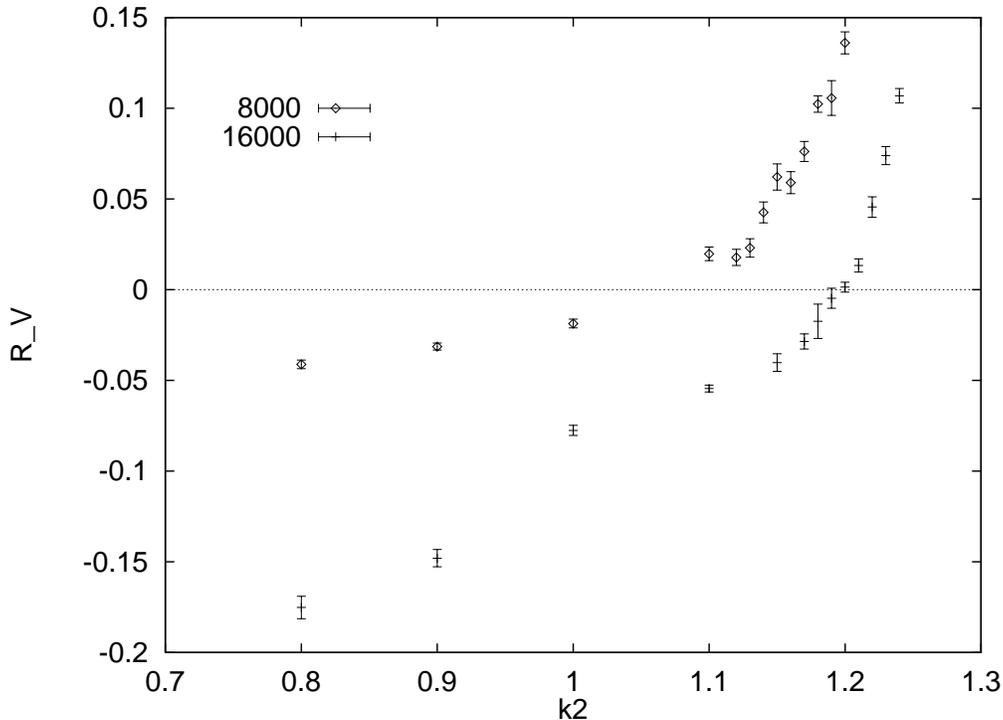}
\caption{The effective curvature $R_V$ as a function of $\kappa_2$ for
  8000 and 16000 simplices.}
\label{curvaturefig}
\end{figure}

{}Figure~\ref{curvaturefig} shows the resulting effective curvature
$R_V$ as a function of $\kappa_2$ for 8000 and 16000 simplices.  For
$N=8000$ the fitting range was $r=1$--9.  We see that $R_V$ starts
negative and then rises with $\kappa_2$, going through zero.  In
contrast, the Regge curvature at the triangles $\langle
\overline{R}\rangle$ is positive for all $\kappa_2$ values in the
figure.

The value of $a$ in (\ref{meanrfit}) varied from about 0.9
at $\kappa_2 = 0.8$ to 0.4 near the transition. These numbers are much
smaller than the flat value of 8.47 in equation (\ref{analytica}),
indicating an effective volume per simplex $V_{\mbox{\scriptsize eff}}
\approx 20-50$, much larger than $V_4\approx 2.3$.  This is at least
partly due to the way we measure distances.  The distances are
measured using paths which can only go along the dual lattice and will
therefore be larger than the shortest paths through the simplicial
complex.

\begin{figure}[t]
\epsfxsize=\textwidth
\epsffile{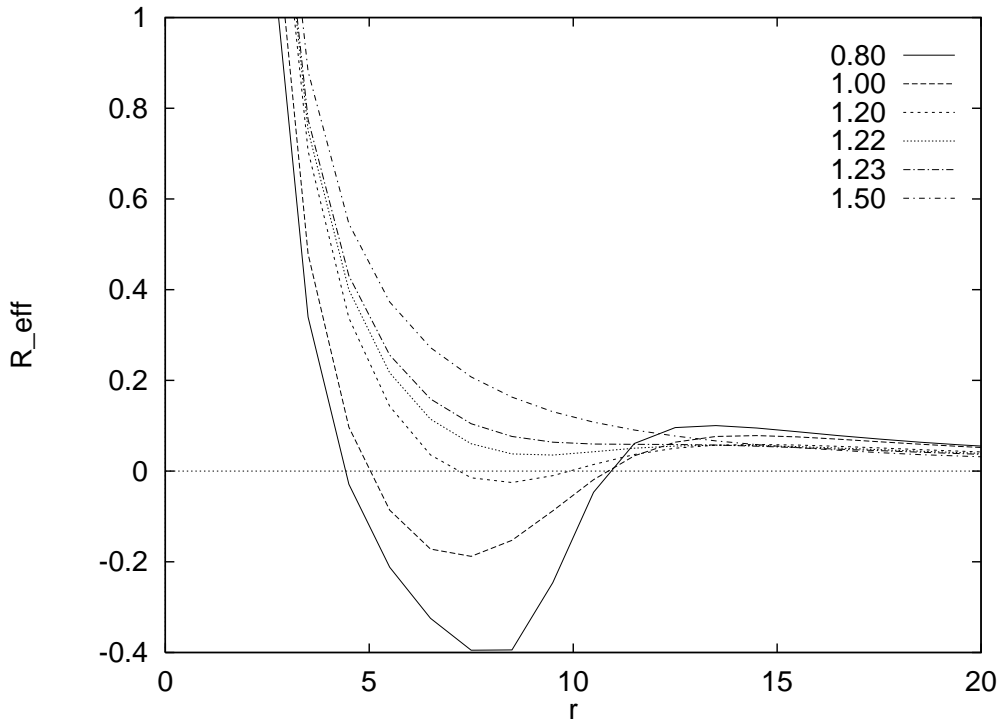}
\caption{The effective curvature $R_{\rm eff}(r)$ for various $\kappa_2$ 
  and 16000 simplices.}
\label{refffig}
\end{figure}

There is a strong systematic dependence of $R_V$ on the range of $r$
used in the fit. For example, for the $N=16000$ data in
figure~\ref{curvaturefig} we used a least squares fit in the range
$r=1$--11.  If the range is changed to $r=1$--9 the data for $R_V$
have to be multiplied with a factor of about 1.5.  We can enhance this
effect by reducing the fitting range to only two $r$ values and
thereby obtain a `running curvature' $R_{\rm eff}(r)$ at distance $r$.
We write $N'(r) = a(r) r^3 + b(r)r^5$, $N'(r+1) = a(r)(r+1)^3 +
b(r)(r+1)^5$ and define $R_{\rm eff}(r+\frac{1}{2}) = -24b(r)/a(r)$,
which gives
\begin{equation}
  R_{\rm eff}(r+{\textstyle\frac{1}{2}}) = 24\frac{(r+1)^3 - r^3
    N'(r+1)/N'(r)} {(r+1)^5 - r^5 N'(r+1)/N'(r)}.
\label{reff}
\end{equation}
{}Figure~\ref{refffig} shows the behavior of $R_{\rm eff}(r)$ for
various $\kappa_2$. It drops rapidly from large values $\approx 4.5$
(which is near $\langle\overline{R}\rangle \approx 5.5$) at $r=0$ to
small values at $r\approx 8$.  For $\kappa_2 \aleq \kappa_2^c$ the
curves have a minimum and around this minimum the values of $R_{\rm
  eff}$ average approximately to the $R_V$'s displayed earlier.

The idea of $R_{\rm eff}(r)$ and $R_V$ is to measure curvature from
the correlation function $N'(r)$ by comparing it with the classical
volume-distance relation for distances $r$ going to zero, as long as
there is reasonable indication for classical behavior at these
distances. Clearly, we cannot let $r$ go to zero all the way because
of the huge increase of $R_{\rm eff}$. This seems to indicate a
`planckian regime' where classical behavior breaks down.

\section{Dimension}
\label{dimension}

One of the interesting observables in the model is the dimension at
large scales.  A common way to define a fractal dimension is by studying the 
behaviour of the volume within a distance $r$ and identifying the 
dimension $d$ if the volume behaves like
\begin{equation}
  V(r) = \mbox{const.} \times r^d.
  \label{simplepower}
\end{equation}
Such a measurement has been done in \cite{AgMi92a,AgMi92b}, using the
geodesic distance and a distance defined in terms of the massive
scalar propagator \cite{Da92}.  Arguments against the necessity of
such use of the massive propagator were raised in ref.\ \cite{Fi92}.
Although we feel that the issue is not yet settled, we shall use here
the geodesic distance as in the previous section.

\begin{figure}[t]
\epsfxsize=\textwidth
\epsffile{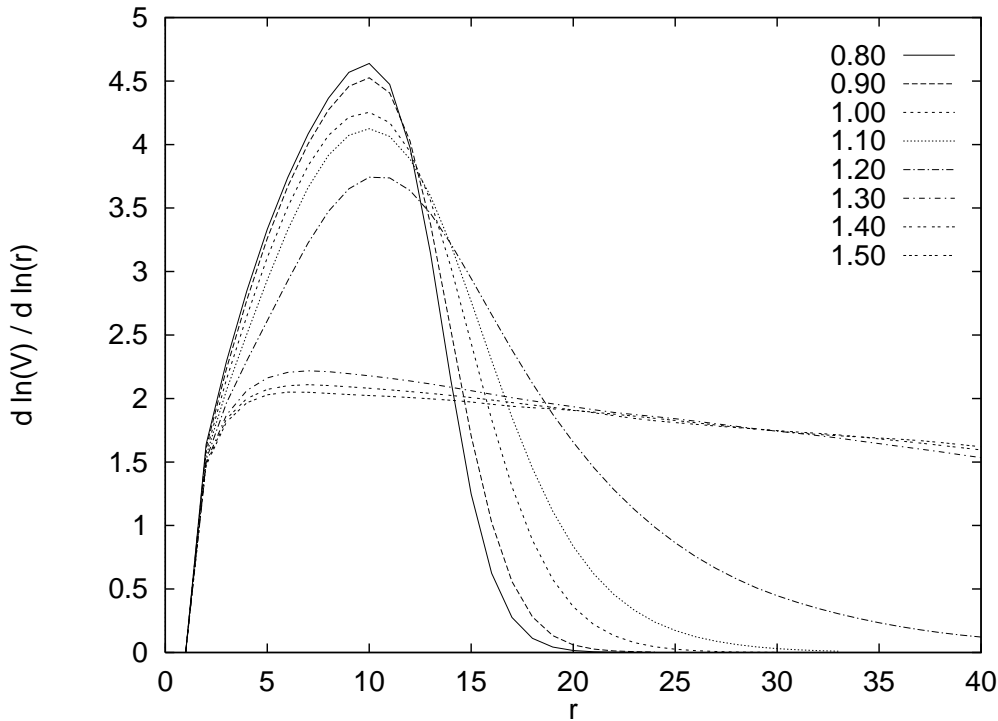}
\caption{$\ln[(N(r)/N(r-1))]/\ln[r/(r-1)] \longleftrightarrow
  d\ln V/d\ln r$ as a function of $r$ for some values of $\kappa_2$.}
\label{dlnvdlnrfig}
\end{figure}

If the volume does go like a power of $r$, the quantity
\begin{equation}
  d = \frac{d\ln V}{d\ln r} \longleftrightarrow \frac{\ln N(r) - \ln
    N(r-1)}{\ln(r)-\ln(r-1)},
\end{equation}
would be a constant.  Figure~\ref{dlnvdlnrfig} shows this quantity for
some values of $\kappa_2$ for a system with 16000 simplices.  We see a
sharp rise at distances $r=1,2$ independent of $\kappa_2$ which is
presumably a lattice artefact. The curves then continue to rise until
a maximum where we may read off an effective dimension, which is
clearly different in the crumpled phase ($\kappa_2 = 0.80$--1.20) and
in the elongated phase ($\kappa_2 = 1.30$--1.50).  Instead of a local
maximum one would of course like a plateau of values where $d\ln
V/d\ln r$ is constant and may be identified with the dimension $d$.
Only for $\kappa_2$ beyond the transition a range of $r$ exists where
$d\ln V/d\ln r$ looks like a plateau.  In this range, $d \approx 2$.

Similar studies have been carried out in 2D dynamical triangulation
where it was found that plateaus only appear to develop for very large
numbers of triangles \cite{2D}.  Our 4D systems are presumably much
too small for $d\ln V/d\ln r$ to shed light on a fractal dimension at
large scales, if it exists.  However, we feel that the approximate
plateau in the elongated phase with $d=2$ should be taken seriously.

As we are studying a system with the topology of the sphere $S^4$, it
seems reasonable to look whether it behaves like a 
$d$-dimensional sphere $S^d$. For such a hypersphere with 
radius $r_0$, the volume behaves like 
\begin{equation}
   V'(r) = d C_d\, r_0^{d-1} (\sin \frac{r}{r_0})^{d-1}.
  \label{effdimdef}
\end{equation}
This prompts us to explore (\ref{effdimdef}) as a definition of the
dimension $d$, we shall call it the global dimension. 
For small $r/r_0$ this reduces to (\ref{simplepower}).
On the other hand for $d=n$ the form (\ref{effdimdef}) is compatible
with the effective curvature form (\ref{vpofcur}), with
\begin{equation}
  R_V = \frac{n(n-1)}{r_0^2}.
\label{currad}
\end{equation}

To determine the dimension $d$ we fit the data for $N'(r)$ to a 
function of the form (\ref{effdimdef}),
\begin{equation}
  N'(r) = c\, (\sin\frac{r}{r_0})^{d-1}.
\label{efdfit}
\end{equation}
The free parameters are $r_0$, $d$ and the multiplicative constant
$c$. It is a priori not clear which distances we need to use to make
the fit. At distances well below the maximum of $N'(r)$ (cf.\ 
figure~\ref{dvofrdrfig}) the effective curvature fit appears to give a
reasonable description of the data, but it will be interesting to try
(\ref{effdimdef}) also for these distances.  Small distances are of
course affected by the discretization.  This is most pronounced at low
$\kappa_2$ where the range of $r$-values is relatively small.  On the
other hand, for small $\kappa_2$ the fits turn out to be good up to
the largest distances, indicating a close resemblance to a
hypersphere, while at larger $\kappa_2$ the values of $N'(r)$ are
asymmetric around the peak (cf.\ fig.~\ref{dvofrdrfig}) and fits turn
out to be good only up to values of $r$ not much larger than this
peak. The system behaves like a hypersphere up to the distance where
$N'(r)$ has its maximum, which would correspond to halfway the maximum
distance for a real hypersphere.  Above that distance it starts to
deviate, except for small $\kappa_2$ where $N'(r)$ is more symmetric
around the peak and the likeness remains.

For $\kappa_2 = 0.8$ (figure~\ref{fitcru}) the global dimension fit
(\ref{efdfit}) was performed to the data at $r=7$--21, for $\kappa_2 =
1.22$ (figure~\ref{fittra}) to $r=4$--20 and for $\kappa_2 = 1.50$
(figure~\ref{fitelo}) to $r=3$--60.

The two descriptions, effective curvature at lower distances and
effective dimension at intermediate and larger distances, appear
compatible.  Notice that at $\kappa_2=0.8$ in the crumpled phase the
local effective curvature $R_V$ is negative while the global structure
resembles closely a (positive curvature) sphere with radius $r_0=7.6$
($r_0= 2r_m /\pi$, with $r_m$ the value where $N'(r)$ is maximal). At
$\kappa_2=1.22$ near the transition the effective curvature and
effective dimension descriptions appear to coincide. At
$\kappa_2=1.50$, deep in the elongated phase, the effective curvature
fit does not make sense anymore, its $r$-region of validity has
apparently shrunk to order 1 or less. The effective dimension fit on
the other hand is still good in this phase and the power behavior
(\ref{simplepower}) with $d\approx 2$ is extended by (\ref{efdfit})
to intermediate distances including the maximum of $N'(r)$.

\begin{figure}[t]
\epsfxsize=\textwidth
\epsffile{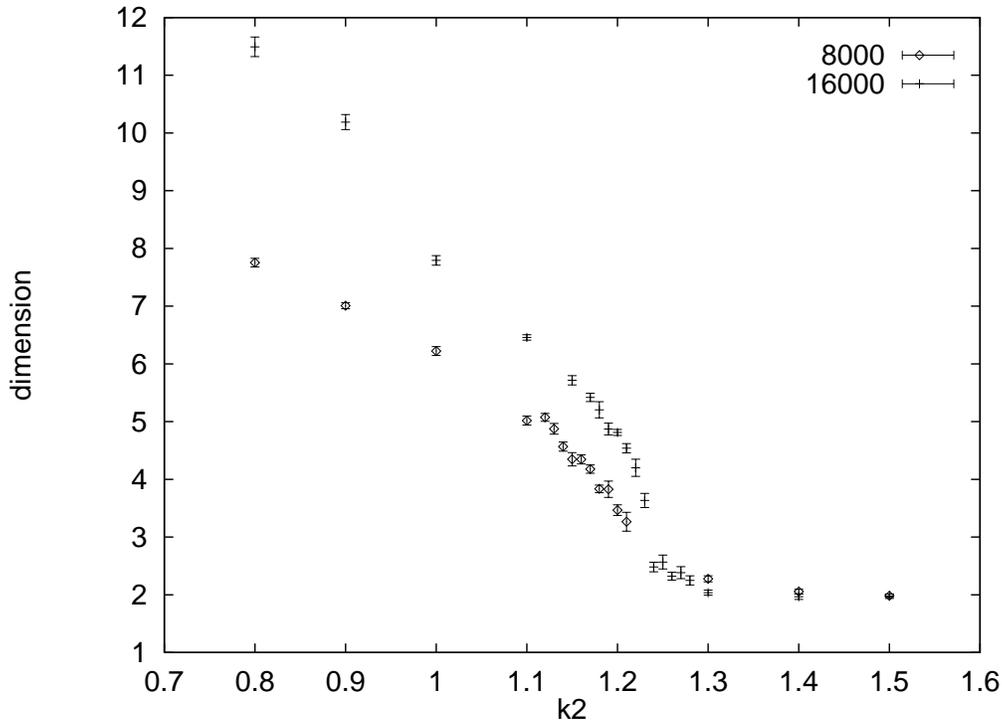}
\caption{The global dimension as a function of $\kappa_2$ for 8000
  and 16000 simplices.}
\label{dimensfig}
\end{figure}

Figure~\ref{dimensfig} shows the global dimension as a function of
$\kappa_2$ for the total volumes of 8000 and 16000 simplices.  For
small values of $\kappa_2$ it is high and increases with larger
volumes.  For values of $\kappa_2$ beyond the transition it quickly
goes to two, confirming the statement made earlier that in this region
$N'(r)$ is approximately linear with $r$ down to small $r$.

A most interesting value of the dimension is the one at the phase
transition.  To determine the value of $\kappa_2^c$ where the
transition takes place we look at the curvature susceptibility of the
system, figure~\ref{suscepfig}.  For 8000 simplices the peak in the
susceptibility is between $\kappa_2$ equals 1.17 and 1.18 where the
dimensions we measured are 4.2(1) and 3.8(1).  For 16000 simplices the
peak is between 1.22 and 1.23 where the dimensions are 4.2(1) and
3.6(1).  Therefore the dimension at the transition is consistent with
4.  As can be seen from these numbers the largest uncertainty in the
dimension is due to the uncertainty in $\kappa_2^c$.  The effective
dimensions have some uncertainty due to the ambiguity of the range of
$r$ used for the fit.  Near the transition this generates an extra
error of approximately 0.1.

\begin{figure}[t]
\epsfxsize=\textwidth
\epsffile{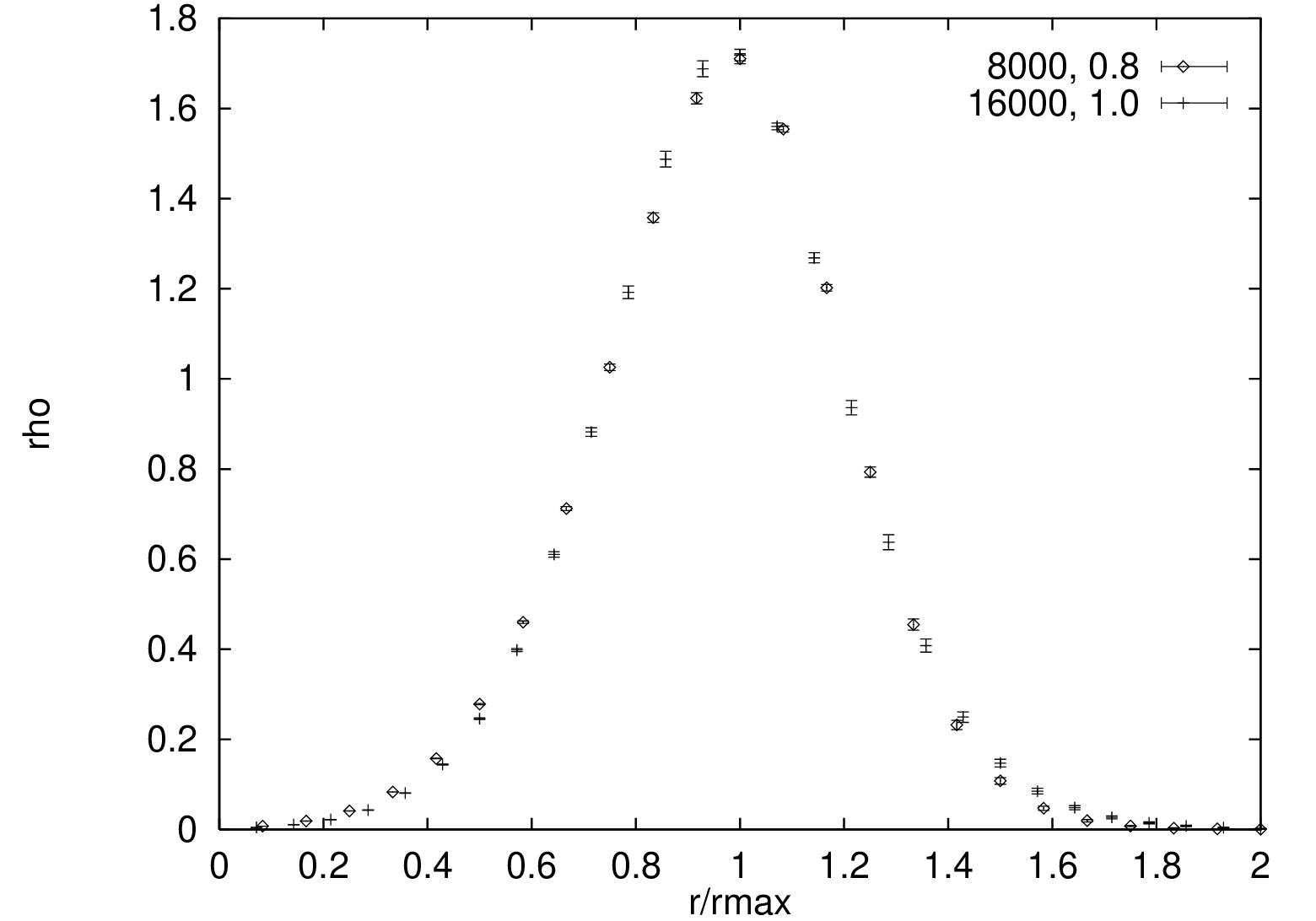}
\caption{The scaling function $\rho$ for $\kappa_2 = 0.8$ at $N =
  8000$ and $\kappa_2 = 1.0$ at $N = 16000$.}
\label{scalcru}
\end{figure}

\begin{figure}[t]
\epsfxsize=\textwidth
\epsffile{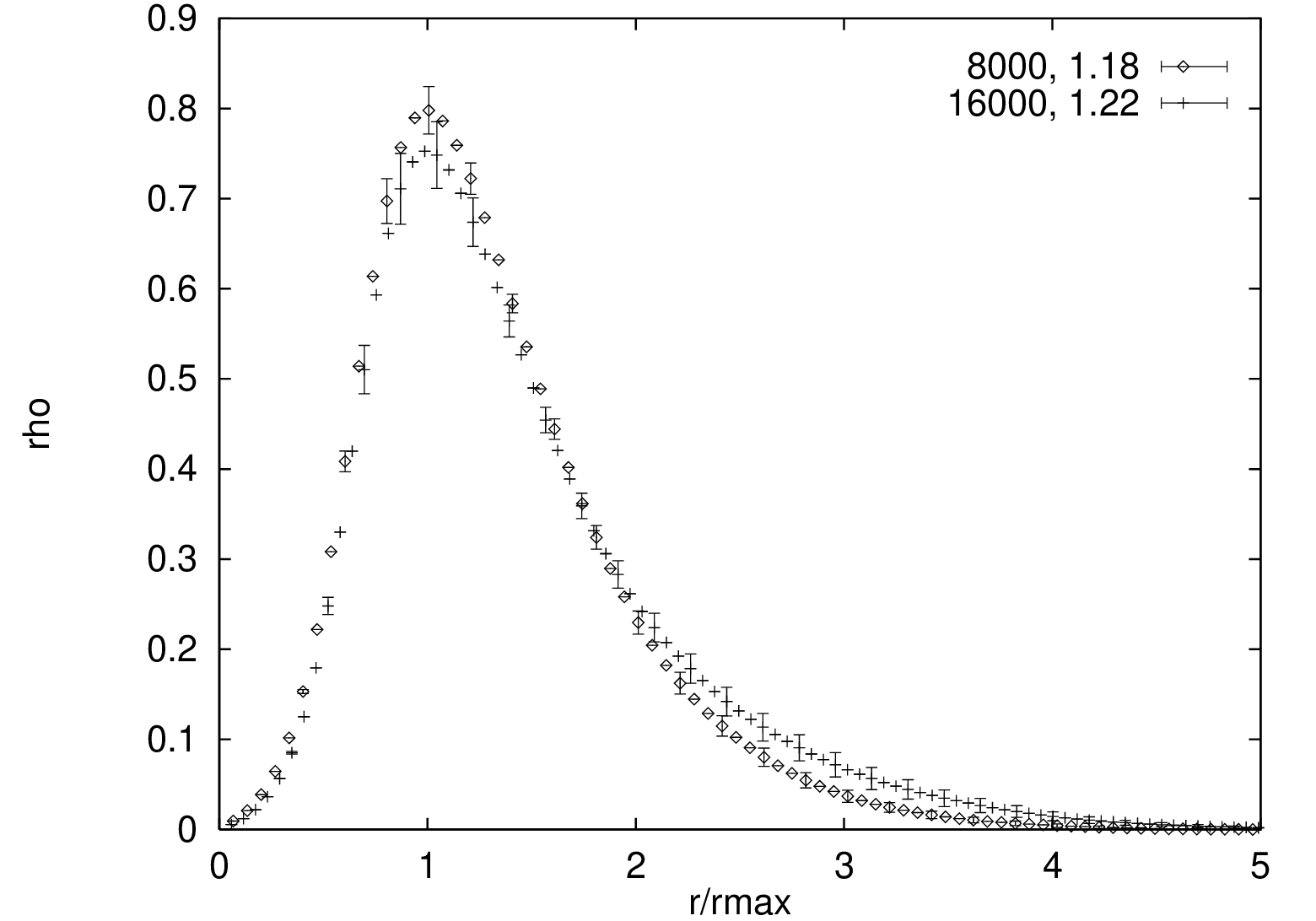}
\caption{The scaling function $\rho$ for $\kappa_2 = 1.18$ at $N =
  8000$ and $\kappa_2 = 1.22$ at $N = 16000$.}
\label{scaltra}
\end{figure}

\begin{figure}[t]
\epsfxsize=\textwidth
\epsffile{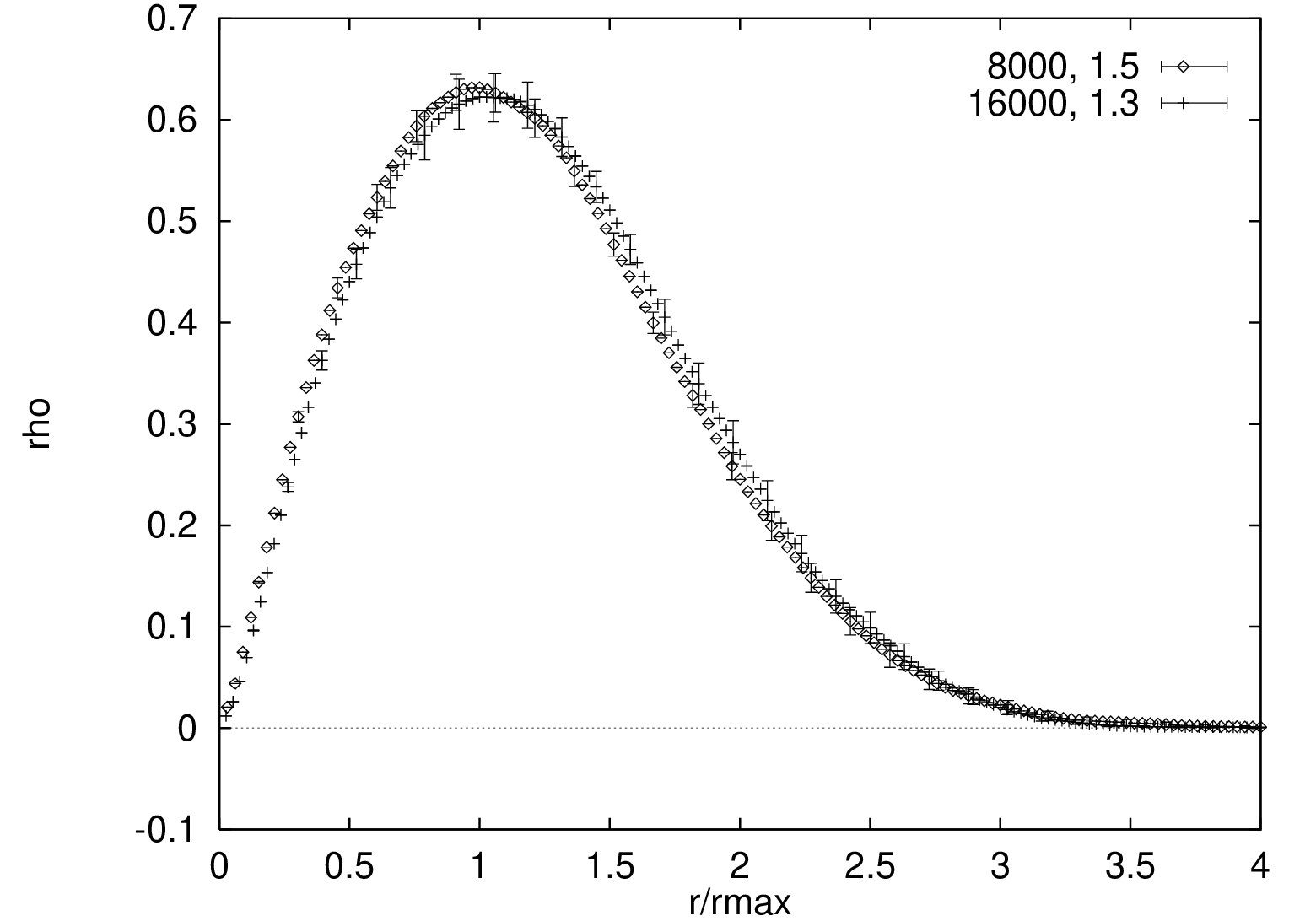}
\caption{The scaling function $\rho$ for $\kappa_2 = 1.5$ at $N =
  8000$ and $\kappa_2 = 1.3$ at $N = 16000$.}
\label{scalelo}
\end{figure}

\section{Scaling}
\label{scaling}

To get a glimpse of continuum behavior it is essential to find scaling
behavior in the system. We found a behavior like a $d$ dimensional
hypersphere for $r$ values up to the value $r_m$ where $N'(r)$ is
maximal. This suggests scaling in the form
\begin{equation}
  N'(r)= r_m^{d-1} f(\frac{r}{r_m},d),
\label{scal1}
\end{equation}
i.e.\ $N'(r)$ depends on $\kappa_2$ and $N$ through $d=d(\kappa_2,N)$
and $r_m = r_m(\kappa_2,N)$. The occurence of $d$ in this formula is
unattractive, however, since it is obtained by comparing $N'(r)$ to
$\sin^{d-1} (r/r_0)$ at intermediate scales, which is a somewhat
imprecise concept. We would like a model independent test of scaling.

Consider the probability for two simplices to have a geodesic
distance $r$,
\begin{equation}
  p(r) = \frac{N'(r)}{N},\;\;\;1=\sum_{r=1}^{\infty} p(r) \approx \int
  dr\, p(r),
\end{equation}
which depends parametrically on $\kappa_2$ and $N$.  It seems natural
to assume scaling for this function in the form
\begin{equation}
  p(r) = \frac{1}{r_m} \rho(\frac{r}{r_m},\tau),\;\;\; \int
  dx\,\rho(x,\tau)=1,
  \label{scal2}
\end{equation}
where $\tau$ is a parameter playing the role of $d$ in (\ref{scal1})
which labels the different functions $\rho$ obtained this way. For
instance, $\tau$ could be the value $\rho_m=\rho(1,\tau)$ at the
maximum of $\rho$. This may give give problems if $\rho_m$ does not
change appreciably with $\kappa_2$ (similar to $d$ in the elongated
phase). Other possibilities are $\tau=\overline{r}p(r_m)$ or $\tau =
\overline{r^k} / \overline{r}^k$ for some $k$ with $\overline{r^k} =
\sum_r p(r) r^k$.  In practice we may also simply take
$\tau=\kappa_2-\kappa_2^c(N)$ at some standard choice of $N$ and
compare the probability functions with the $p(r)$ at this $N$.

Matched pairs of $\rho(x,\tau)$ for $N=8000$ and 16000 are shown in
figures~\ref{scalcru}--\ref{scalelo}, respectively far in the crumpled
phase, near the transition and in the elongated phase.  For clarity we
have left out part of the errors. Scaling appears to hold even for
$\kappa_2$ values we considered far away from the transition.

The values of $\kappa_2(N)$ of the matched pairs in
figs.~\ref{scalcru}--\ref{scalelo} are increasing with $N$ in the
crumpled phase and decreasing with $N$ in the elongated phase.  This
suggests convergence from both sides to $\kappa_2^c(N)$ as $N$
increases. For current system sizes $\kappa_2^c(N)$ is still very much
dependent on $N$, a power extrapolation estimate \cite{CaKoRe94a}
gives $\kappa_2^c(\infty)\approx 1.45$.

We can use the matched pairs of $\rho$ to define a scaling dimension
$d_s$ by
\begin{equation}
  N = \alpha\, r_m^{d_s},
  \label{scal4}
\end{equation}
where $\alpha$ and $d_s$ depend only on $\tau$.  Using non-lattice
units, replacing the integer $r_m$ by $r_m/(\ell/\sqrt{10})$ where
$r_m$ is now momentarily dimensionful, we can interpret (\ref{scal4})
as
\begin{equation}
  N \propto \left(\frac{r_m}{\ell}\right)^{d_s},
  \label{scal4a}
\end{equation}
which shows that the scaling dimension characterizes the
dimensionality of the system at small scales $\ell\rightarrow 0$ with
$r_m$ fixed. Returning to lattice units, taking for $r_m$ the integer
value of $r$ where $N'(r)$ is maximal and using an assumed error of
0.5, these scaling dimensions would be 4.5(3), 3.8(2) and 4.0(1)
respectively for figures~\ref{scalcru},~\ref{scaltra}
and~\ref{scalelo}.  The largest errors in these figures probably
arises due to the uncertainty in the values of $\kappa_2$ we need to
take to get matching curves, i.e.\ to get the same value of $\tau$.
As we do not have data for a continuous range of $\kappa_2$ values, we
have to make do with what seems to match best from the values we do
have.  Nevertheless, the values of $d_s$ far away from the transition
are strikingly close to 4 when compared to the values of the global
dimension $d$, which are 7.8 and 2.0 for figures~\ref{scalcru}
and~\ref{scalelo}.

The scaling form (\ref{scal1}) is in general incompatible with
(\ref{scal2}), except for $d_s=d$. The evidence for $d_s=4$ points
instead to a scaling behavior of the form
\begin{equation}
  N'(r) = r_m^3 f(\frac{r}{r_m},\tau),
\end{equation}
with $f(x,\tau) = \alpha({\tau})\rho(x,\tau)$. This further suggests
scaling of the volume $V'(r)$ at distance $r$ with an effective volume
$V_{\mbox{\scriptsize eff}}$ per simplex (cf.\ (\ref{VN})) depending
only on $\tau$.

A precise definition of the scaling form $\rho(x,\tau)$ may be 
given by
\begin{equation}
  \mbox{$\rho(x,\tau)= r_m p(r_m x)$,\ $\kappa_2=\kappa_2(N)$ such that
    $r_m p(r_m)=\tau$,\ $N\rightarrow\infty$},
  \label{scal3}
\end{equation}
where we used $\tau=\rho_m$ for illustration.
Intuitively one would expect the convergence to the scaling limit 
(\ref{scal3}) to be non-uniform, with the large $x$ region 
converging first, and there may be physical aspects to such nonuniformity. 

\begin{figure}[t]
\epsfxsize=\textwidth
\epsffile{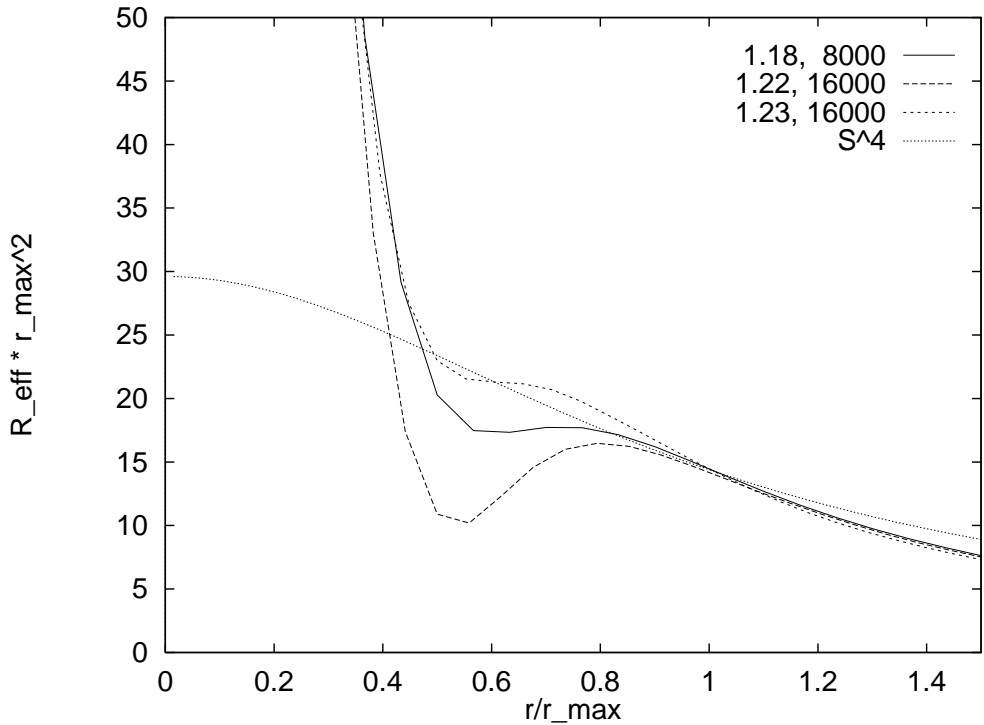}
\caption{Scaling form $r_m^2 R_{\rm eff}(xr_m)$ versus $r/r_m$ near the 
  the transition, for $N=8000$, $\kappa_2=1.18$ (middle) and
  $N=16000$, $\kappa_2 =1.22$, 1.23 (lower and upper). The
  hypothetical limiting form corresponding to $S^4$ is also shown.}
\label{refftra}
\end{figure}

The scaling analogue of running curvature (\ref{reff}) is given by
\begin{equation}
  \tilde{R}(x) \equiv r_m^2\, R_{\rm eff} (xr_m) = \frac{24}{x^2} \,
  \frac{3-d\ln\rho/d\ln x}{5 - d\ln\rho/d\ln x}.
\end{equation}
Fig.~\ref{refftra} shows this function for the matched pair of
fig.~\ref{scaltra} in the transition region.  We have also included
the curve for $\kappa_2=1.23$ at $N=16000$.  The curves do not match
in the region around $x=0.5$ and $R_{\rm eff}$ is apparently a
sensitive quantity for scaling tests.  Still, fig.~\ref{refftra}
suggests reasonable matching for a value of $\kappa_2(16000)$
somewhere between 1.22 and 1.23 and even the steep rise as far as
shown appears to be scaling approximately, with $\tilde{R}(16000)$
somewhat below $\tilde{R}(8000)$.  We find similar scaling behavior
for the matched pair in the elongated phase.  The number of our
$\kappa_2$ values in the crumpled phase is too limited to be able to
draw a conclusion there.

The steep rise appears to move to the left for increasing $N$ (a
scaling violation).  A most interesting question is whether the onset
of the rise (e.g.\ the $x$ value where $\tilde{R}=50$) continues to
move towards $x=0$ as $N \to \infty$. Such behavior is needed for a
classical region to open up from $x$ around 1 towards the origin
$x=0$. In other words, the `planckian regime' would have to shrink in
units of the size $r_m$ of the `universe'. Then $r_m^2 R_V$ could be
defined as a limiting value of $\tilde{R}$, $x \to 0$.

Since we are looking for classical behavior in the transition region
it is instructive to compare with the classical form of $\tilde{R}$ 
corresponding to the sphere $S^4$, for which
$\rho = \rho_m \sin^3\theta$, $\theta=\pi x/2$, 
\begin{equation}
  \tilde{R} = \frac{18\pi^2}{\theta^2}\,\frac{\tan\theta - \theta}
  {5\tan\theta - 3\theta} = 3\pi^2 (1- \frac{13\pi^2}{120}\,x^2 +
  \cdots).
\label{rtilclas}
\end{equation}
This function is also shown in fig.~\ref{refftra}.
Our current data are evidently still far from the hypothetical
classical limiting form (\ref{rtilclas}).

\section{Summary}
\label{summary}

The average number of simplices $N'(r)$ at geodesic distance $r$ gives
us some basic information on the ensemble of euclidean spacetimes
described by the partition function (\ref{partn}). The function
$N'(r)$ is maximal at $r=r_m$ and $r_m N'(r)/N$ shows scaling when
plotted as a function of $r/r_m$. We explored a classical definition
of curvature in the small to intermediate distance regime based on
spacetime dimension four, the effective curvature $R_V$. We also
explored a description at global distances by comparing $N'(r)$ with a
sphere of effective dimension $d$. Judged by eye, the effective
curvature fits and effective dimension fits give a reasonable
description of $N'(r)$ in an appropriate distance regime
(figs.~\ref{fitcru}--\ref{fitelo}).  The resulting $R_V$ depends
strongly on the fitting range, which led us to an explicitly distance
dependent quantity, the `running' curvature $R_{\rm eff}(r)$. This
dropped rapidly from lattice values of order of the Regge curvature at
$r=0$ to scaling values of order of the $R_V$ found in the effective
curvature fits at intermediate distances.  A preliminary analysis of
scaling behavior then suggested the possibility of a classical regime
with a precise definition of $r_m^2 R_V$ in the limit of large $N$.
We shall now summarize the results further, keeping in mind the
ambiguity in $R_V$ as derived from the fits to $N'(r)$.

For small $\kappa_2$ the effective curvature is negative.  Furthermore
the system resembles a $d$-sphere with very large dimension $d$ which
increases with the volume.  This suggests that no matter how large the
volumes we use, there will never be a region of $r$ where the power
law $V(r)\propto r^d$ gives good fits over large ranges of $r$.  In
other words, the curves for $d\ln V/d\ln r$ in
figure~\ref{dlnvdlnrfig} will never have a plateau.  This behavior is
consistent with that of a space with constant negative curvature,
where the volume rises exponentially with the geodesic distance for
distances larger than the curvature radius and if we look at large
enough scales the intrinsic fractal dimension equals infinity. The
resulting euclidean spacetime cannot be completely described as a
space with constant negative curvature as such a space with topology
$S^4$ does not exist, and finite size effects take over at still
larger distances.

At the transition the spacetime resembles a 4-dimensional sphere with
small positive effective curvature, up to intermediate distances.

For large $\kappa_2$ the system has dimension 2.  In this region it
appears to behave like a branched polymer, which has an intrinsic
fractal dimension of 2 \cite{Da92}.  Moving away from the transition,
the curvature changes much more rapidly than in the small $\kappa_2$
phase and the effective curvature radius $r_V$ soon becomes of the
order of the lattice distance. A priori, two outcomes seem plausible.
In the first, the system collapses and $r_V$ becomes of order 1 in
lattice units, reflecting the unboundedness of the continuum action
{}from below.  In the second, the spacetime remains 4-dimensional and
$r_V$ can still be tuned to large values in lattice units, but very
small compared to the global size $r_m$, for a sufficiently large
system.  So far the second outcome seems to be favored, for two
reasons.  Firstly, the function $N'(r)$ looks convex for $r \leq 6$
slightly above the transition, e.g.\ for $\kappa_2 = 1.24$ at 16000
simplices. In other words, the linear behavior as seen in
figure~\ref{fitelo} does not set in immediately above the transition.
Secondly, the system shows scaling and the scaling dimension $d_s$ as
defined in (\ref{scal4}) is approximately 4 even far into the
elongated phase, indicating 4-dimensional behavior at small scales.

High and low dimensions in the crumpled and elongated phases with the
value four at the transition were reported earlier in
ref.~\cite{AgMi92b}.  This dimension was apparently interpreted as a
small scale dimension, whereas instead we find a small scale dimension
of four in all phases.  Similar results are also found in the Regge
calculus approach to quantum gravity \cite{hamber1,markum1}, where one
also finds two phases, a strong (bare Newton) coupling phase with
negative curvature and fractal dimension four, and (using an $R^2$
term in the action for stabilization) a weak coupling phase with
fractal dimension around two.

\section{Discussion}
\label{discussion}

The evidence for scaling indicates continuum behavior. This 
brings up a number of issues which need to be addressed in a 
physical interpetation of the model.

One guideline in this work is the question whether there is a regime
of distances where $V'(r) = V_{\rm eff} N'(r)$ behaves classically for
suitable bare Newton constant $G_0$. The connection with classical
spacetime can be strengthened by identifying the geodesic distance $r$
with a cosmic time $t$ and $V'(r)^{1/3}$ with the scale factor $a(t)$
in a euclidean Robertson-Walker metric. The classical action for
$a(r)$ is given by
\begin{equation}
  S = -\frac{\pi}{8 G}\int dr\, [a \left(\frac{da}{dr}\right)^2 + a] +
  \lambda\left(2\pi^2\int dr\, a^3 -V\right),
\label{msp}
\end{equation}
where $\lambda$ is a Lagrange multiplier enforcing a total spacetime
volume $V$. It plays the role of a cosmological constant which is just
right for getting volume $V$.  For positive $G\lambda$ the solution of
the equations of motion following from (\ref{msp}) is $a=r_0\sin
r/r_0$, which represents $S^4$ with $R=12/r_0^2=192 \pi G \lambda$.
For negative $G\lambda$ the solution is $a=r_0\sinh r/r_0$ which
represents a space of constant negative curvature $R=-12/r_0^2=192\pi
G\lambda$, cut off at a maximal radius to get total volume $V$.

The form (\ref{msp}) serves as a crude effective action for the system
for intermediate distances around the maximum in $V'(r)$ at $r_m$, and
couplings $\kappa_2\propto G_0^{-1}$ in the transition region. At
larger distances the fluctuations of the spacetimes grow, causing
large baby universes and branching, and averaging over these may be
the reason for the asymmetric shape of $V'(r)$. Because of this the
Robertson-Walker metric cannot give a good description at these
distances.

Intuitively one expects also strong deviations of classical behavior
at distances of order of the Planck length $\sqrt{G}$, assuming that a
Planck length exists in the model. A proposal for measuring it was put
forward in \cite{BaSm94a}. The steep rise in the running curvature
$R_{\rm eff}(r)$ at smaller $r$ indeed suggests such a planckian
regime. It extends to rather large $r$ but it appears to shrink
compared to $r_m$ as the lattice distance decreases, i.e. $N$
increases for a given scaling curve labelled by $\tau$.  This suggests
that the Planck length goes to zero with the lattice spacing, $G /
r_m^2\to 0$ as $N\to \infty$ at fixed $\tau$.  This does not
necessarily mean that the Planck is length of order of the lattice
spacing, although this is of course quite possible.  However, the
theory may also scale at planckian distances and belong to a
universality class.  It might then be `trivial'.

At this point it is instructive to recall other notorious models with
a dimensionful coupling as in Einstein gravity, the 4D nonlinear sigma
models. The lattice models have been well studied, in particular the
$O(4)$ model for low energy pion physics (see for example
\cite{LATO4}).  It has one free parameter $\kappa=\ell^2 f_0^2$ which
corresponds to the renormalized dimensionful coupling $f^2$; $f$ is
the pion decay constant or the electroweak scale in the application to
the Standard Model. With this one bare parameter it is possible to
tune {\em two\/} quantities, $f/m$ and $\ell f$, where $m$ is the mass
of the sigma particle or the Higgs particle.  This trick is possible
because the precise value of $\ell$ is unimportant, as long as it is
sufficiently small\footnote{It is good to keep the numbers in
  perspective: for example in the Standard Model $f=250$ GeV and for a
  Higgs mass $m= 100$ GeV or less, $\ell$ is 15 orders of magnitude
  smaller than the Planck length, or even much smaller.}.  However, in
the continuum limit $\ell f\rightarrow 0$ triviality takes its toll:
$m/f\rightarrow 0$ and the model becomes noninteracting.  The analogy
$f^2 \leftrightarrow 1/G$, $m \leftrightarrow r_m^{-1}$ suggests that
we may be lucky and there is a scaling region in $\kappa_2$--$N$
space, for a given scaling curve (given $\tau$), where the theory has
universal properties and where we can tune $G/r_m^2$ to a whole range
of desired values.  Taking the scaling {\em limit\/} however might
lead to a trivial theory with $G/r_m^2 =0$. In case this scenario
fails it is of course possible to introduce more parameters, e.g.\ as
in $R^2$ gravity, to get more freedom in the value of $G/r_m^2$. This
then raises the question of universality at the Planck scale.

We really would like to replace $r_m^2$ by $R_V$ in the reasoning in
the previous paragraph, since we view $R_V$ as the local classical
curvature, provided that a classical regime indeed develops as $N \to
\infty$.

Next we discuss the nature of the elongated phase. Even deep in this
phase we found evidence for scaling. Furthermore, for given scaling
curve, increasing $N$ means decreasing $\kappa_2$. Hence, increasing
$N$ at fixed $\kappa_2$ brings the system deeper in the elongated
phase. This leads to the conclusion that there is nothing wrong with
the elongated phase. It decribes very large spacetimes which are two
dimensional on the scale of $r_m$ but not necessarily at much smaller
scales. It could be effectively classical at scales much larger than
the Planck length but much smaller than $r_m$.

This reasoning further suggests that $N$ and $\kappa_2$ primarily
serve to specify the `shape' of the spacetime. The tuning $\kappa_2
\approx \kappa_2^c$ is apparently not needed for obtaining criticality
but for obtaining a type of spacetime. The peak in the susceptibility
of the Regge curvature could be very much a reflection of shape
dependence.  Most importantly, this suggests that the physical
properties associated with general coordinate invariance will be
recovered automatically in 4D dynamical triangulation, as in 2D with
fixed topology\footnote{A field theory analogue is $Z_n$ lattice gauge
  theory, which for $n \geq 5$ has been found to posses a Coulomb
  phase, a whole region in bare parameter space with massless photons;
  see for example ref.\ \cite{ZN}.}.

\section*{Acknowledgements}
This work is supported in part by the Stichting voor Fundamenteel
Onderzoek der Materie (FOM).  The numerical simulations were partially
carried out on the IBM SP1 at SARA.


\begin{thebibliography}{99}

\bibitem{AgMi92a} M.E.~Agishtein and A.A.~Migdal, Mod. Phys. Lett. A7
  (1992) 1039.

\bibitem{AgMi92b} M.E.~Agishtein and A.A.~Migdal, Nucl. Phys. B385
  (1992) 395.

\bibitem{AmJu92} J.~Ambj\o rn and J.~Jurkiewicz, Phys. Lett. B278
  (1992) 42.

\bibitem{We82} D.~Weingarten, Nucl. Phys. B210 (1982) 229.

\bibitem{AmJuKr93} J.~Ambj\o rn, J.~Jurkiewicz and C.F.~Kristjansen,
  Nucl. Phys. B393 (1993) 601.

\bibitem{Var93} S.~Varsted, Nucl. Phys. B412 (1994) 406.

\bibitem{Br93} B.~Br\"ugmann, Phys. Rev. D47 (1993) 3330.

\bibitem{CaKoRe94a} S.~Catterall, J.~Kogut and R.~Renken, Phys. Lett.
  B28B (1994) 277.

\bibitem{AmJaJuKr93} J.~Ambj\o rn, S.~Jain, J.~Jurkiewicz and
  C.F.~Kristjansen, Phys. Lett. B305 (1993) 208.

\bibitem{CaKoRe94b} S.~Catterall, J.~Kogut and R.~Renken, Phys. Rev.
  Lett. 72 (1994) 4062.

\bibitem{AmJu94} J.~Ambj\o rn and J.~Jurkiewicz {\it On the
    exponential bound in four dimensional simplicial gravity},
  preprint NBI-HE-94-29.

\bibitem{BruMa94} B.~Br\"ugmann and E.~Marinari, {\it More on the
    exponential bound of four dimensional simplicial quantum gravity},
  MPI-PHT-94-72.

\bibitem{BaSm94b} B.V.~de~Bakker and J.~Smit, Phys. Lett. B334 (1994)
  304.

\bibitem{Da92} F.~David, Nucl. Phys. B368 (1992) 671.

\bibitem{2D} M.E.~Agishtein and A.A.~Migdal, Nucl. Phys. B350 (1991)
  690; N.K.\penalty200 Kawamoto, V.A.~Kazakov, Y.~Saeki and
  Y.~Watabiki, Phys. Rev. Lett. 68 (1992) 2113; J.~Ambj\o rn, P.~Bia\l
  as, Z.~Burda, J.~Jurkiewicz and B.~Petersson, Phys. Lett. B342
  (1995) 58.

\bibitem{Fi92} T.~Filk, Mod. Phys. Lett. A7 (1992) 2637.

\bibitem{hamber1} H.W.~Hamber, Nucl. Phys. B400 (1993) 347.

\bibitem{markum1} W.~Beirl, E.~Gerstenmayer, H.~Markum and J.~Riedler,
  Phys. Rev. D49 (1994) 5231.

\bibitem{BaSm94a} B.V.~de Bakker and J.~Smit, Nucl. Phys. B (Proc. 
  Suppl.) 34 (1994) 739.

\bibitem{LATO4} M.~L\"uscher and P.~Weisz, Phys. Lett. B212 (1988)
  472; Nucl. Phys. B318 (1989) 705; U.M.~Heller, Nucl. Phys. 
  (Proc. Suppl.) 34 (1994) 101.

\bibitem{ZN} J.~Fr\"olich and T.~Spencer, Comm. Math. Phys. 83
  (1982) 411; V. Alessandrini, Nucl. Phys. B215 (1983) 337;
  V.~Alessandrini and Ph.~Boucaud, Nucl. Phys. B225 (1983) 303.

\end{thebibliography}
\end{document}